\documentclass[%
aps,
prapplied,
superscriptaddress,
longbibliography,
 amsmath,amssymb,
 reprint,%
]{revtex4-2}

\usepackage{hyperref}

\usepackage[capitalise]{cleveref}

\usepackage{graphicx}% Include figure files
\usepackage{dcolumn}% Align table columns on decimal point
\usepackage{bm}% bold math
\usepackage{float}
\usepackage{array}
\usepackage{textcomp}
\usepackage{physics}
\usepackage[dvipsnames]{xcolor}
\usepackage{multirow}
\usepackage{booktabs}
\usepackage{upgreek}
\usepackage{xr}
\usepackage{cleveref}
\usepackage{tabularx}
\usepackage{amsmath}
\usepackage{appendix}
\usepackage{gensymb}
\usepackage{fixltx2e}

\usepackage{newtxtext}
\usepackage{newtxmath}

\usepackage[mathlines]{lineno}% Enable numbering of text and display math
\newcolumntype{C}[1]{>{\centering\arraybackslash}p{#1}}

\begin{document}

\preprint{AIP/123-QED}

\title[]{Precision frequency tuning of tunable transmon qubits using alternating-bias assisted annealing}

\author{Xiqiao Wang}
\email{xwang@rigetti.com}
\affiliation{ 
Rigetti Computing, 775 Heinz Avenue, Berkeley, CA 94710, USA
}
\author{Joel Howard}
\affiliation{ 
Rigetti Computing, 775 Heinz Avenue, Berkeley, CA 94710, USA
}
\author{Eyob A. Sete}
\affiliation{ 
Rigetti Computing, 775 Heinz Avenue, Berkeley, CA 94710, USA
}
\author{Greg Stiehl}
\affiliation{ 
Rigetti Computing, 775 Heinz Avenue, Berkeley, CA 94710, USA
}
\author{Cameron Kopas}
\affiliation{ 
Rigetti Computing, 775 Heinz Avenue, Berkeley, CA 94710, USA
}
\author{Stefano Poletto}
\affiliation{ 
Rigetti Computing, 775 Heinz Avenue, Berkeley, CA 94710, USA
}%
\author{Xian Wu}
\affiliation{ 
Rigetti Computing, 775 Heinz Avenue, Berkeley, CA 94710, USA
}
\author{Mark Field}
\affiliation{ 
Rigetti Computing, 775 Heinz Avenue, Berkeley, CA 94710, USA 
}%
\author{Nicholas Sharac}
\affiliation{ 
Rigetti Computing, 775 Heinz Avenue, Berkeley, CA 94710, USA
}
\author{Christopher Eckberg}
\affiliation{ 
Rigetti Computing, 775 Heinz Avenue, Berkeley, CA 94710, USA
}
\author{Hilal Cansizoglu}
\affiliation{ 
Rigetti Computing, 775 Heinz Avenue, Berkeley, CA 94710, USA
}
\author{Raja Katta} 
\affiliation{ 
Rigetti Computing, 775 Heinz Avenue, Berkeley, CA 94710, USA
}
\author{Josh Mutus}
\affiliation{ 
Rigetti Computing, 775 Heinz Avenue, Berkeley, CA 94710, USA
}
\author{Andrew Bestwick}
\affiliation{ 
Rigetti Computing, 775 Heinz Avenue, Berkeley, CA 94710, USA
}
\author{Kameshwar Yadavalli}
\affiliation{ 
Rigetti Computing, 775 Heinz Avenue, Berkeley, CA 94710, USA
}
\author{David P. Pappas}
\affiliation{ 
Rigetti Computing, 775 Heinz Avenue, Berkeley, CA 94710, USA
}

\date{May 1, 2024}
%\date{\today}

\begin{abstract}
Superconducting quantum processors are one of the leading platforms for realizing scalable fault-tolerant quantum computation (FTQC). The recent demonstration of post-fabrication tuning of Josephson junctions using alternating-bias assisted annealing (ABAA) technique and a reduction in junction loss after ABAA illuminates a promising path towards precision tuning of qubit frequency while maintaining high coherence. Here, we demonstrate precision tuning of the maximum $|0\rangle\rightarrow |1\rangle$ transition frequency ($f_{01}^{\rm max}$) of tunable transmon qubits by performing ABAA at room temperature using commercially available test equipment. We characterize the impact of junction relaxation and aging on resistance spread after tuning, and demonstrate a frequency equivalent tuning precision of 7.7 MHz ($0.17\%$) based on targeted resistance tuning on hundreds of qubits, with a resistance tuning range up to $18.5\%$. Cryogenic measurements on tuned and untuned qubits show evidence of improved coherence after ABAA with no significant impact on tunability. Despite a small global offset, we show an empirical $f_{01}^{\rm max}$ tuning precision of 18.4 MHz by tuning a set of multi-qubit processors targeting their designed Hamiltonians. We experimentally characterize high-fidelity parametric resonance iSWAP gates on two ABAA-tuned 9-qubit processors with fidelity as high as $99.51\pm 0.20\%$. On the best-performing device, we measured across the device a median fidelity of $99.22\%$ and an average fidelity of $99.13\pm 0.12 \%$. Yield modeling analysis predicts high detuning-edge-yield using ABAA beyond the 1000-qubit scale. These results demonstrate the cutting-edge capability of frequency targeting using ABAA and open up a new avenue to systematically improving Hamiltonian targeting and optimization for scaling high-performance superconducting quantum processors. 
\end{abstract}

% Valid PACS numbers may be entered using the \verb+\pacs{#1}+ command.
%Use showkeys class option if keyword

\maketitle

%introduction
\section{Introduction}
Scaling of fault-tolerant quantum computing using superconducting quantum information processors requires two-qubit gates with fidelities $99\%$ or better across the device \cite{PhysRevA.80.052312}. Among major classes of qubit architectures, square lattices of tunable qubits with a tunable coupler between each pair of qubits represent a promising architecture for their ability to provide isolation for single-qubit gates while enabling fast two-qubit gate control with dynamic coupling~
\cite{Yan2017, arute2019a, Stehlik2021, Sung2021, Mundada2019, Sete_float_2021, Sete_para_2021,Field2023}. A key challenge for scaling tunable coupling lattice architecture is precise tuning of the tunable qubit’s maximum $|0\rangle\rightarrow |1\rangle$ transition frequency ($f_{01}^{\rm max}$, the flux sweet spot) to the designed Hamiltonian pattern where qubit-qubit detuning is carefully optimized to within a tight distribution range to (1) allow activating a target gate (iSWAP or CZ) on each two-qubit edge without parking the qubits off of $f_{01}^{\rm max}$, (2) enable shallow flux modulation to suppress pulse distortion and leakage, and preserve qubit coherence under modulation, and (3) avoid frequency collision between nearest and next-nearest-neighbor qubits while parking qubits at $f_{01}^{\rm max}$\cite{Sete_para_2021}. However, typical frequency variations in the as-fabricated transmon qubits are limited to a few percent due to challenges in controlling junction critical dimensions in the nanometer scale \cite{Kreikebaum_2020,osman2021simplified, pishchimova2023improving}. Previously, laser annealing \cite{granata2007localized, Hertzberg2021, LasiQScience2022, LasiQBerkeley2022}, thermal annealing \cite{koppinen2007complete, migacz2003thermal, korshakov2024aluminum}, and e-beam annealing \cite{balaji2024electronbeam} on fixed-frequency transmon qubits have been reported for frequency tuning, among which laser annealing demonstrated the state-of-the-art frequency tuning precision of 18.5 MHz (empirical) and 4.7 MHz (frequency-equivalent resistance precision) with no measurable impact on qubit coherence. Recently, alternating bias assisted annealing (ABAA) of amorphous oxide tunnel junctions \cite{pappas2024alternating} emerged as a promising, alternative technique for scalable precision frequency tuning offering a unique combination of advantages, which includes reduction of junction loss and two-level defects in Josephson junction barriers, a large resistance tuning range ($>$ 70\% at $80^{\circ} C$), and ease of implementation requiring only conventional test equipment. However, precise frequency tuning and Hamiltonian targeting using ABAA have not been reported till now.

In this paper, we adapt the ABAA technique for post-fabrication trimming of $f_{01}^{\rm max}$ to the desired Hamiltonian patterns and demonstrate high-fidelity two-qubit gates on ABAA-tuned multi-qubit quantum processors. We characterize the impact of junction relaxation and aging on resistance spread following ABAA tuning for resistance targeting calibration. Combining baseline statistics of targeted resistance tuning on hundreds of qubits and cryogenic frequency prediction characterization, we demonstrate a frequency-equivalent resistance tuning precision of 7.7 MHz, which corresponds to a frequency spread of 0.17\% around the predicted average qubit frequency. On cooling down both ABAA-tuned and untuned tunable transmon qubits, we observe evidence of improved coherence after ABAA frequency trimming with no measurable impact on tunability. We tune a set of multi-qubit processors targeting their designed Hamiltonians and show an empirical frequency tuning precision of 18.4 MHz in $f_{01}^{\rm max}$, including additional deviations from pre-cooldown processes, frequency prediction, and $f_{01}^{\rm max}$ measurement conditions. We experimentally demonstrate high-fidelity parametric resonance two-qubit iSWAP gates on two ABAA-tuned 9-qubit processors with fidelity as high as $99.51 \pm 0.20\%$. We calibrated iSWAP gates on all edges of both devices achieving a median fidelity (gate time) of $99.00\%$ (68 ns) and $99.22\%$ (80 ns), respectively. We discuss the implications of the improved $f_{01}^{\rm max}$ targeting precision to detuning edge yield and scaling. We envision that the ABAA technique will be broadly applied as an easy-to-implement, high-precision frequency tuning approach for scaling high-performance superconducting quantum processors.  

\begin{figure}[t]
\includegraphics[width=\columnwidth]{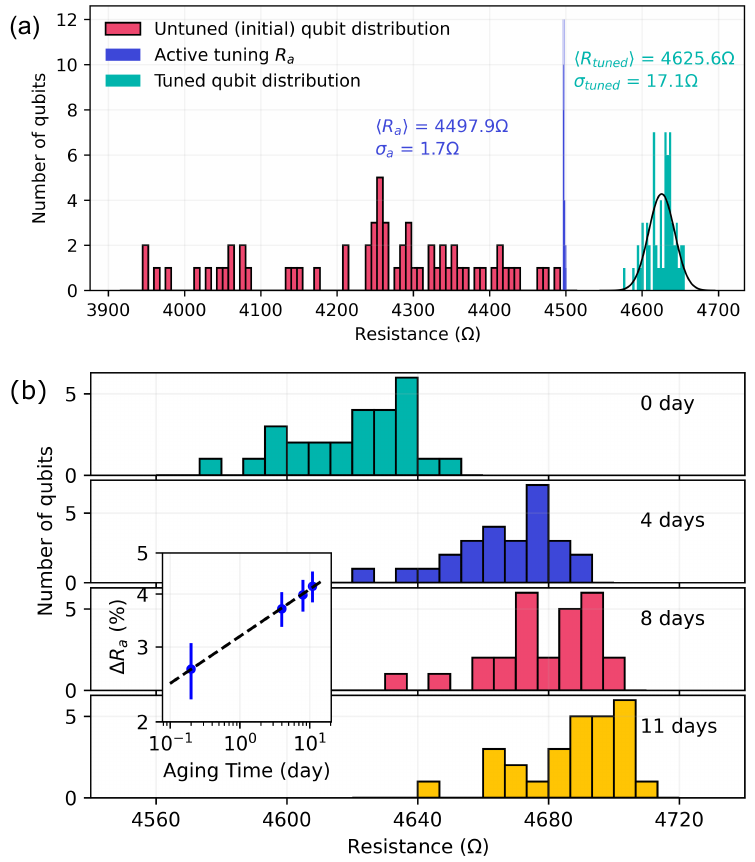}
\caption{Relaxation and aging after active ABAA tuning. (a) Resistance distributions of the same set of qubits at different stages of resistance tuning. The initial distribution (pink, $7.5 \Omega$ bin width) is tuned by ABAA on a qubit-by-qubit basis to the blue distribution ($0.5 \Omega$ bin width), which consists of the resistance measured at each qubit once a pre-set threshold ($4496 \Omega$) is surpassed, which also triggers the termination of the active tuning pulse sequence. The blue distribution then evolves into the distribution in cyan ($3 \Omega$ bin width) by the time when ABAA tuning is completed on all components and the tuned resistances are probed within the same day. (b) Time evolution of the tuned resistance distribution ($6.7 \Omega$ bin width) monitored on a subset of the population shown in (a) at 4 days, 8 days, and 11 days after the tuned resistance is first measured on day 0. Inset: a power-law fit ($y=3.18\times x^{0.11}$) to the time evolution of the mean values of the tuned distribution. Error bars represent one standard deviation of the distributions.}
\label{Fig:Figure1}
\end{figure}

\begin{figure}[t]
\includegraphics[width=\columnwidth]{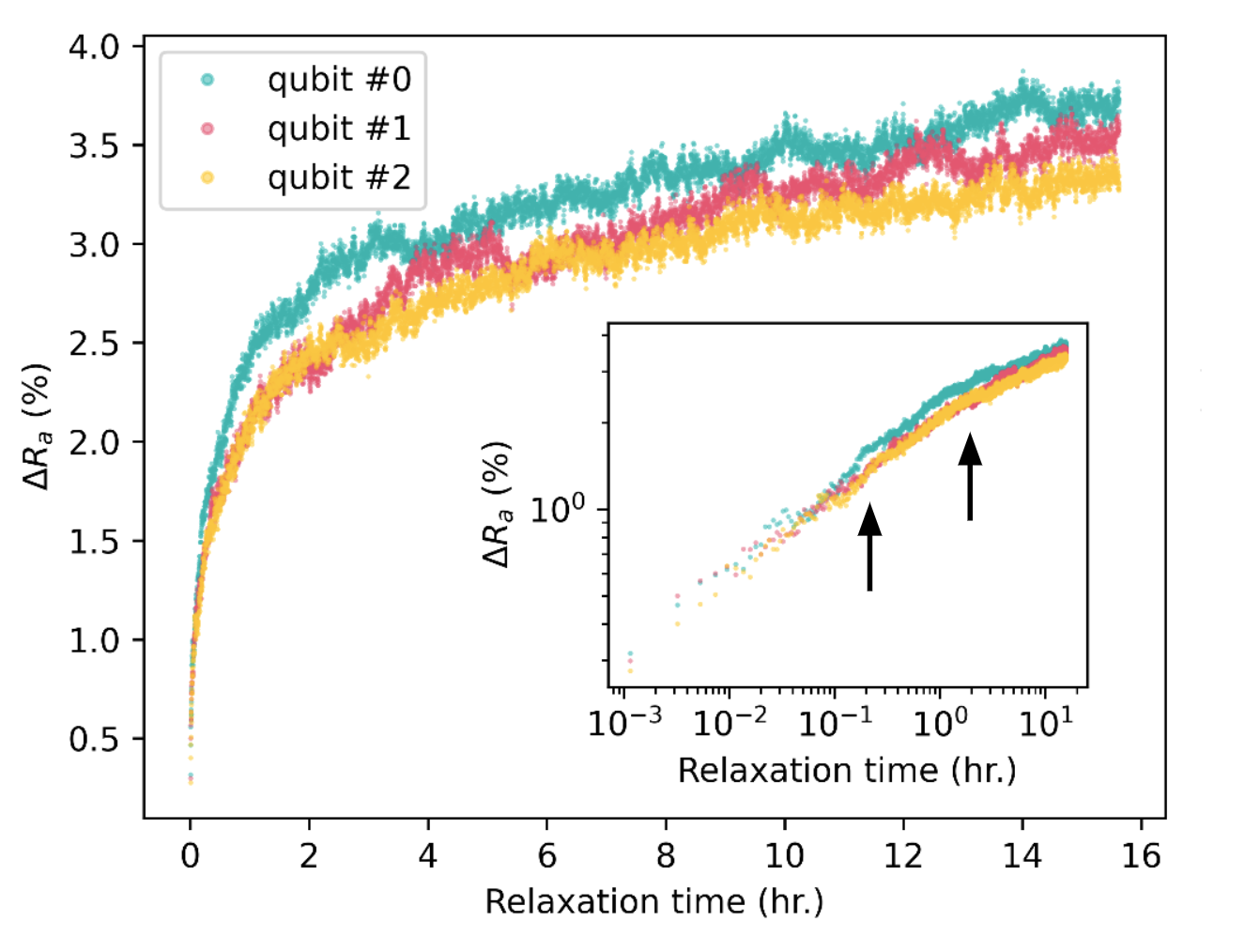}
\caption{Change in resistance monitored on three qubits during junction relaxation following ABAA tuning to the same threshold resistance ($7711 \ \Omega$). Inset: same dataset plotted in log-log scale. Arrows in the inset mark the 2 distinct kinks that approximately divide the relaxation curves into short ($ t<0.2\ \mathrm{hr}$), intermediate ($ 0.2<t<2\ \mathrm{hr}$), and long ($ t>2\ \mathrm{hr}$) sections. The slopes in the log-log plot can be characterized by the exponent of power-law fit ($y=\beta x^{\alpha}$ with $\beta$ and $\alpha$ the fitting parameters). The fitted exponents ($\alpha$) for the three qubits are 0.30$\pm0.02$, 0.24$\pm0.01$, and 0.16$\pm0.01$ for short, medium and long times, respectively.
}
\label{Fig:Figure2_v4}
\end{figure}

\section{Relaxation and aging after ABAA}
In this work, we perform ABAA adaptively to incrementally adjust the qubit resistance. An ABAA pulse sequence \cite{pappas2024alternating} is applied simultaneously to both Josephson Junctions (JJ) within a tunable transmon qubit. Qubit resistance is monitored at each ABAA tuning pulse cycle to determine if a pre-determined threshold resistance has been reached. The sample is kept at room temperature during the entire tuning operation to avoid parasitic conductance through the substrate for precise resistance measurement, and to suppress Josephson Junction (JJ) aging and resistance drifting during active ABAA tuning \cite{korshakov2024aluminum}. A typical ABAA tuning workflow starts with probing the as-fabricated resistance ($R_{\rm untuned}$) of all qubits on a quantum processor. The fabrication target is set to be below the designed target and within the suitable range for ABAA tuning. The tuning target resistance $R_{T}$ is determined from the designed Hamiltonian of a qubit lattice using a previously calibrated frequency prediction relation, also reserving a global aging budget (typically $2\%$) in anticipation of continued aging after ABAA tuning until the cooldown test. ABAA tuning is then performed sequentially on a qubit-by-qubit basis till all qubits are tuned. Finally, all qubits on the processor are re-probed, denoting as $R_{\rm tuned}$, and compared with the tuning target for evaluating the resistance tuning precision.

To assess the impact of junction relaxation and aging behavior on resistance tuning, as shown in Fig. 1(a), we tune a set of qubits (61 in total) to the same threshold resistance of $4496\ \Omega$. For each qubit, we stop the ABAA pulsing sequence immediately after detecting the last monitor resistance $R_{a}$ surpassing the $4496\ \Omega$ threshold and move on to tune the next qubit. After tuning a batch, we measure the distribution of the tuned resistance $R_{\rm tuned}$. The average waiting time between the last tuning pulse on each component and the time when $R_{\rm tuned}$ is probed is approximately 5 hours for the tuned population shown in Fig. 1(a). The distributions of $R_{\rm untuned}$, monitor resistance $R_{a}$ at the last tuning pulse, and $R_{\rm tuned}$ are plotted in pink, blue, and cyan, respectively. $R_{a}$ centers on $1.9\ \Omega$ above the $4496\ \Omega$ threshold with a standard deviation of $1.7\ \Omega$, representing an ultimate tuning precision set by the finite tuning step size. The distribution in $R_{\rm tuned}$ exhibits a $127.7 \ \Omega$ shift to the right of the blue distribution with a standard deviation of $17.1\ \Omega$. As will be discussed in the following section, this global shift in mean value results from the relaxation behavior within JJ barrier, and the increase in spread is dominated by the slight difference in relaxation at each component by the time when $R_{\rm tuned}$ is probed.

A subset of ABAA-tuned qubits (28 in total) from Fig. 1(a) is monitored during a prolonged aging period covering the typical period between fabrication and a cryogenic cooldown, up to 11 days after ABAA tuning [see Fig. 1 (b)]. The global mean shift in the aging of the tuned distribution slows down quickly with a power-law exponent fit of 0.11, as shown in the inset log-log plot. However, the distribution spread does not increase over time. This is consistent with the expectation from the exponential slow-down of glassy relaxation \cite{GlassyJunctionsPhysRevB} and barrier oxidation \cite{Cabrera_1949} processes that most relaxation/aging occurs within the first few hours and the rate of change quickly approaches zero afterward. Meanwhile, witness components (untuned) from the same wafer over the same period do not exhibit appreciable aging, attributing the observed global aging shift in Fig. 1(b) to ABAA tuning.

We examined the temporal behavior of resistance relaxation following the ABAA pulse sequence to probe the inherent uncertainty associated with targeting precision and to identify potential factors contributing to variability in tuned resistance levels. Figure 2 illustrates our continuous monitoring of resistance over 15 hours immediately following the final tuning pulse for three ABAA-tuned qubits. Even though these three qubits were tuned to the same threshold resistance $R_{a} = 7711\ \Omega$, they settled to slightly different values with similar time dependencies over time. When plotted on a log-log scale, inset in Fig. 2, it can be seen that they exhibit linear behavior with slightly different time constants over short, intermediate, and long time scales. This indicates that there may be different relaxation mechanisms that depend strongly on the nano-structure of the junctions. While identification of the specific mechanisms in these windows of time is out of the scope of this study, possible mechanisms in the Al and junction include 1) stress and relaxation in the junction in the bulk $\mathrm{AlO}_{x}$ or at the $\mathrm{Al/AlO}_{x}$ interfaces \cite{Al2O3GlassTransitionTHashimoto2022, GlassyJunctionsPhysRevB}; and 2) oxygen diffusion back and forth between the junction and electrodes into the bulk Al or grain boundaries \cite{schafer1991annealing, OxidationReview, OxidationKinetics2012}. Further work to investigate these effects, for example, using functional TEM, modifying the junction interfaces and aluminum grain structure, and applying stress to the junctions, may help elucidate the origin of the spread in relaxation.

\begin{figure*}
\includegraphics[width=\textwidth]{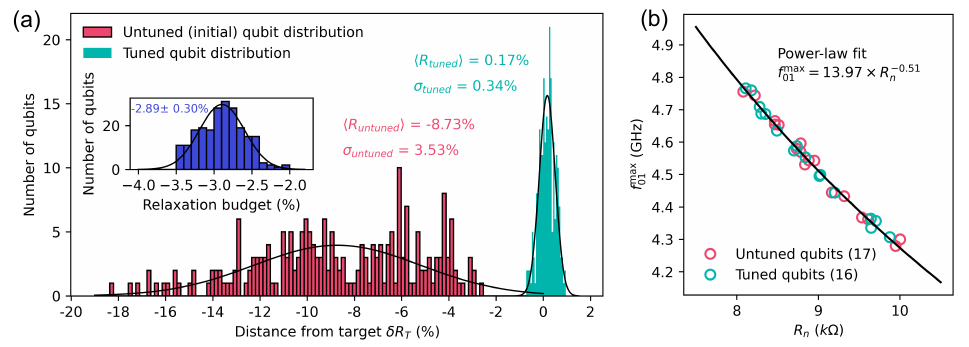}
\caption{Statistics on targeted resistance tuning precision and its frequency-equivalent precision. (a) Resistance distributions of a set of 221 qubits before and after ABAA tuning. The x-axis represents the normalized distance from the measured resistance to the post-tuning target resistance, which is set to $98 \%$ of the design resistance values. The distribution in pink ($0.16 \%$ bin width) and the distribution in cyan (0.06\% bin width) represent, respectively, untuned (as-fabricated) and tuned resistance distributions. The inset plots the distribution of the percentage difference between the resistance measured at the last trimming pulse and the resulting tuned resistance ($R_{\rm tuned}$), characterizing the distribution of the relaxation distances by the time when $R_{\rm tuned}$ is measured. (b) Experimentally measured qubit frequency ($f_{01}^{\rm max}$) versus room-temperature resistance on a set of tuned and untuned tunable transmon qubits. The room-temperature resistance for both tuned and untuned qubits is probed right before packaging (3 days before the cryogenic cooldown date). The black line represents an empirical power-law fit to all data points. }
\label{Fig:Figure3_v4}
\end{figure*}

\section{ABAA tuning precision}
To evaluate the resistance tuning precision for targeting the resistance values assigned based on design Hamiltonians, as shown in Fig. 3 (a), we perform baseline statistical analysis on a set of 221 tuned qubits whose tuning target ($R_{T}$) is set to 98\% of their design values, reserving a pre-cooldown aging budget of 2\%. Normalized to the target resistance on the x-axis in Fig. 3 (a), the untuned (as-fabricated) resistance distribution (pink) is on average 8.7\% below the target (intentional under-targeting in fabrication for post-fabrication tuning) with a spread (1-$\sigma$) of 3.5\%, a spread that is typical to the state-of-the-art JJ fabrication variation \cite{Kreikebaum_2020}. After ABAA tuning, the tuned resistance is approximately normally distributed with a mean value of 0.17\% above the target and a spread (1-$\sigma$) of 0.34\%, showing a $\sim$10-fold reduction in spread from the untuned distribution. We demonstrate a tuning range of up to 18.5\% with this sample set, which is sufficient to cover the typical range of qubit fabrication variations. 

When performing targeted ABAA tuning at each qubit, we reserve a pre-calibrated percentage for junction relaxation and determine the threshold resistance upon which to stop the ABAA pulse sequence at each component. The inset in Fig. 3 (a) shows the distribution of the measured resistance $R_{a}$ at the last tuning pulse with respect to $R_{\rm tuned}$, which corresponds to the relaxation percentage experienced by qubits after the last ABAA tuning pulse till when the tuned resistance is probed. This distribution in $R_{a}$ effectively represents a calibration on when to terminate the ABAA pulse sequence for the resistance to evolve through relaxation to the target resistance $R_{T}$. We find the relaxation carries the resistance by an average distance of 2.89\% with a standard deviation of 0.30\% before reaching $R_{\rm tuned}$, with no appreciable dependence on the total amount of ABAA tuning and the initial or final resistance values during the active ABAA tuning. The close agreement between the precision in $R_{\rm tuned}$ (0.34\% against the tuning target) and the variation in relaxation (0.3\%) highlights the junction-dependent relaxation behavior as a dominant factor to resistance tuning imprecision in this study. 

To correlate the resistance tuning precision characterized at room temperature to its equivalent frequency tuning precision of $f_{01}^{\rm max}$ at cryogenic temperatures, we experimentally measure the $f_{01}^{\rm max}$ of a set of tuned and untuned tunable transmon qubits and extract the frequency prediction relation using an empirical power law fit, as shown in Fig. 3(b). We found that both the tuned and untuned qubits fit well to a single power-law curve, indicating no appreciable impact from ABAA tuning on the frequency prediction used for assigning target resistance. The fitted exponent is $\sim 0.51$, in good agreement with the 0.5 exponent expected from the transmon theory and Ambegaokar-Baratoff relation \cite{koch2007charge, PhysRevLett.10.486}. The standard deviation of the fitting residues is 12.4 MHz, representing an empirical frequency prediction imprecision that includes potential variations in the superconducting gap as well as resistance deviations arising from additional pre-cooldown processes, such as chip cleaning and packaging \cite{LasiQScience2022}. Combining the resistance tuning precision (0.34\%) from Fig. 3 (a) and the empirical power law fit from Fig. 3(b), we derive the frequency equivalence of the ABAA resistance tuning precision to be $\sigma_{f}=\delta f_{01}^{\rm max}/\delta R \cdot \sigma_{R} = 7.7\ \mathrm{MHz}$ (0.17\% of the predicted mean frequency 4556MHz), which is comparable to the state-of-the-art equivalent frequency tuning precision (4.7MHz) demonstrated by IBM using laser annealing on fixed frequency qubits. \cite{LasiQScience2022} 

\begin{figure}
\includegraphics[width=\columnwidth]{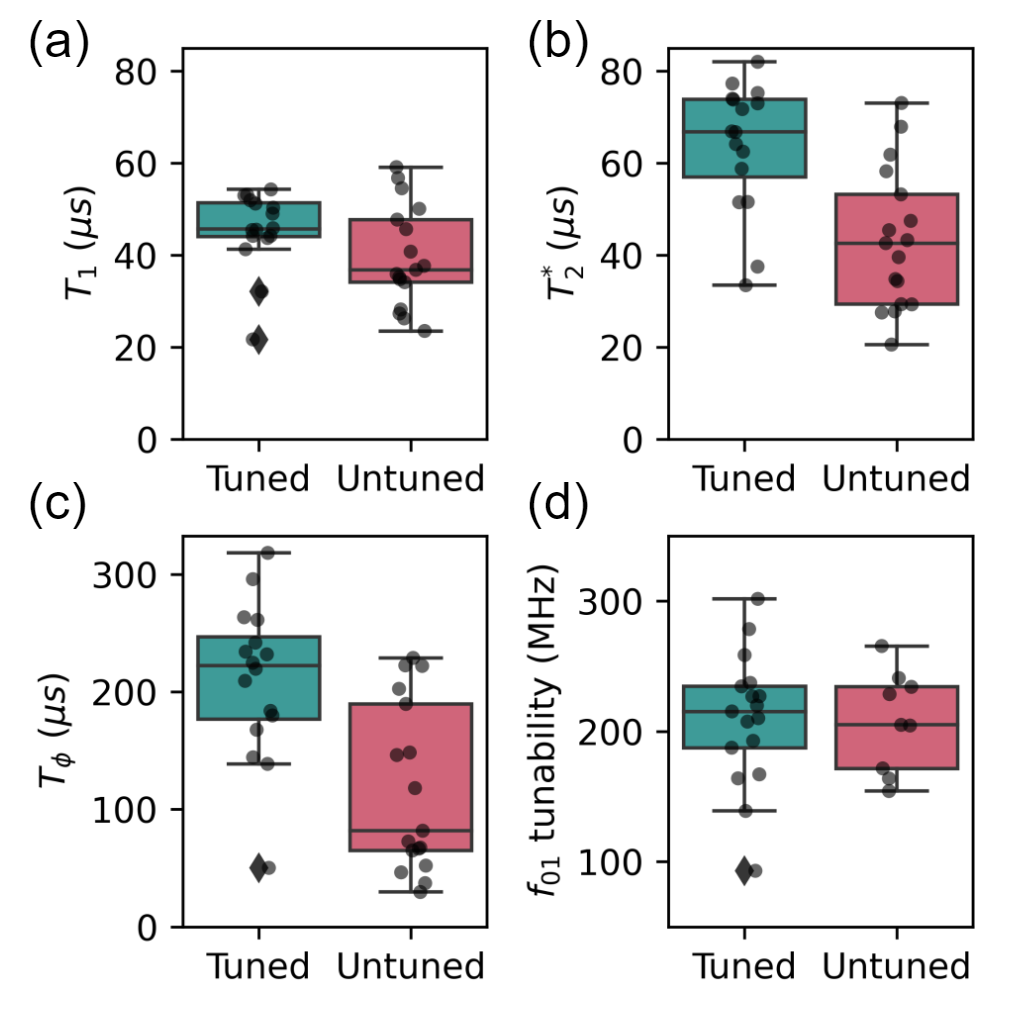}
\caption{Impact of ABAA tuning on the coherence and tunability of tunable transmon qubits. The first three panels show box plots comparing (a) qubit relaxation time $T_{1}$, (b) qubit Ramsey decoherence time $T_{2}^{*}$, and (c) qubit pure dephasing time $T_{\phi}$ on tuned and untuned qubits from the same wafer. All coherence data points were measured at $f_{01}^{\rm max}$. (d) shows box plots comparing qubit tunability on tuned and untuned qubits with nominally identical tunability designs from the same wafer (a different wafer from the one for the coherence comparison). The tunability ($f_{01}^{\rm max}-f_{01}^{\rm min}$) is calculated from measured $f_{01}^{\rm max}$ and $f_{01}^{\rm min}$ values at each qubit. }
\label{Fig:Figure4_v5}
\end{figure}

\section{ Qubit coherence and tunability}

Aside from precision frequency tuning to the designed Hamiltonian patterns, maintaining high coherence and desired tunability range is essential for achieving high 2-qubit gate fidelity. To evaluate the impact of ABAA tuning on coherence, we perform the coherence comparison using a coherence test vehicle chip design where qubits are coupled to frequency-multiplexed feedlines through readout resonators without tunable couplers in between nearest neighbor qubits. We measured the qubit relaxation time $T_{1}$, qubit Ramsey decoherence time $T_{2}^{*}$, and qubit pure dephasing time $T_{\phi}$ while parking at $f_{01}^{\rm max}$ on a set of tuned and untuned tunable transmon qubits (from the same wafer) and plot them side-by-side for comparison in Fig. 4 (a, b, c). While the improvement in $T_{2}^{*}$ and $T_{\phi}$ is statistically significant, the improvement in $T_{1}$ is less obvious within the error range. Nevertheless, ABAA-tuned qubits show higher median coherence values across all three coherence indicators. We calculate the frequency normalized loss tangent ($1/(T_{1}\cdot 2\pi f_{01}^{\rm max})$), and observe a reduction in median loss from 9.8e-7 on untuned qubits to 7.4e-7 on tuned qubits, which is consistent with the loss reduction that we have previously observed on ABAA tuned qubits at elevated temperatures \cite{pappas2024alternating}. 

To empirically determine the impact of ABAA on the tunability of tunable transmon qubits, we experimentally measure the $f_{01}^{\rm max}$ and $f_{01}^{\rm min}$ values of a set of ABAA-tuned and untuned qubits with identical tunability design and from a same wafer, and calculate the tunability ($f_{01}^{\rm max}-f_{01}^{\rm min}$) of each qubit. See box plots in Fig. 4 (d). The tunability variations of untuned qubits are dominated by fabrication variation of the two JJs in each tunable transmon qubit. Previous ABAA study on single junctions \cite{pappas2024alternating} has shown that the junction size dependence has relatively small (few \%) correction on ABAA tuning rate over nearly an order of magnitude in junction area, however, as shown in Fig. 4 (d), the difference in tunability between ABAA-tuned qubits and untuned qubits is statistically insignificant within the statistical error of fabrication variation. This result serves as an empirical demonstration that, within an integrated process flow, the ABAA tuning introduces no statistically significant change in tunability. 

\begin{figure}[t]
\includegraphics[width=\columnwidth]{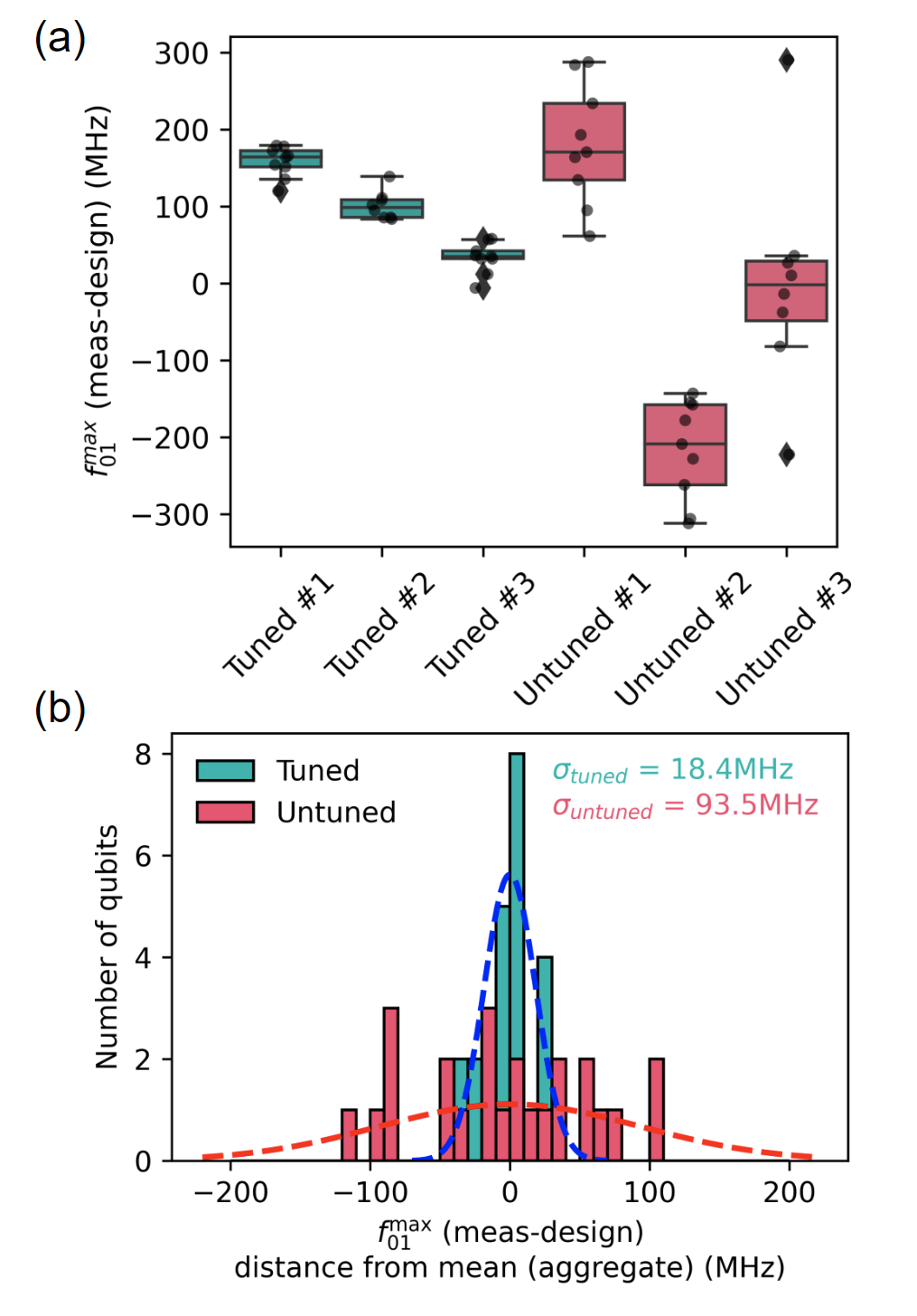}
\caption{Frequency tuning precision on 9-qubit processors. (a) Box plots of the distance from the measured $f_{01}^{\rm max}$ to the design values. Pink boxes represent data from untuned (as-fabricated) processors. Cyan boxes represent data from ABAA-tuned processors. (b) Statistical aggregates of data from (a) after subtracting the global offset on each processor. (bin size = 10 MHz, dashed lines are Gaussian fit) 
}
\label{Fig:Figure5_v4}
\end{figure}

\begin{figure*}
\includegraphics[width=\textwidth]{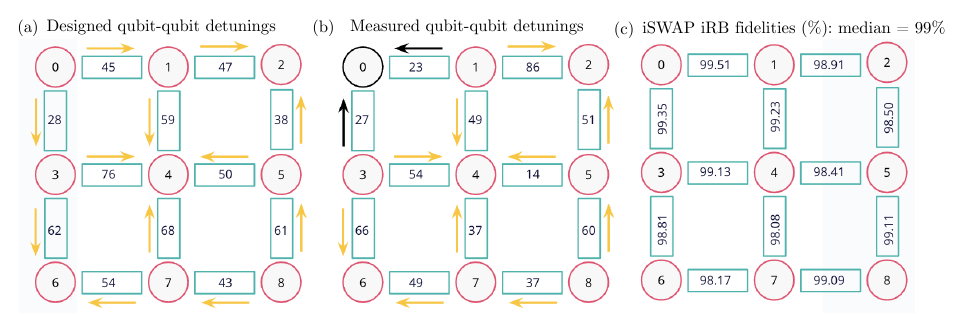}
\caption{Two-qubit gate performance on an ABAA-tuned 9-qubit device. (a) Designed qubit-qubit detunings in $f_{01}^{\rm max}$ between nearest neighbors. (b) Measured qubit-qubit detunings (at $g=0$ tunable coupler flux bias condition, where g is the qubit-qubit coupling) that are used for two-qubit gate operations in (c). Only qubit-0 is parked off of $f_{01}^{\rm max}$ (by -27 MHz) to arrive at the detunings shown in (b), where detunings at all edges fall within the desired small detuning range. All detunings in (a) and (b) are in MHz. The yellow arrows show the directions of the modulated qubits being moved to enact the iSWAP gate. For example, qubit 3 has a higher frequency than qubits 4 and 6 and thus iSWAP gates are activated by moving its frequency down to bring it in resonance with qubits 4 or 6. As shown in (b), the direction of the gate operations changed only on two edges [black arrows in (b)] from (a). (c) Edge distribution of the iSWAP interleaved randomized benchmarking (iRB) fidelity for detuning conditions as shown in (b), reaching median iRB fidelity of $99.00\%$ on the 9-qubit device. The median iSWAP gate time is 68 ns. }
\label{Fig:Figure6_v2}
\end{figure*}

\section{Hamiltonian targeting and two-qubit gate performance}

To demonstrate ABAA’s frequency tuning capability in an integrated workflow for improving detuning edge yield and achieving high two-qubit gate fidelity on full lattices, we trim the tunable qubits on a set of 9-qubit processors targeting the designed Hamiltonian frequency. Qubits on each 9-qubit processor are arranged in a $3\times 3$ square lattice with a tunable coupler between each pair of neighboring qubits. Figure 5(a) plots the deviation from the measured $f_{01}^{\rm max}$ to the design values on three tuned (cyan) and three untuned (pink) processors. The typical measurement uncertainty in $f_{01}^{\rm max}$ is $\sim$4 MHz due to variations in flux bias conditions at adjacent tunable couplers when operating within normal parameter regimes. Despite chip-specific global offsets that are common on both tuned and untuned processors, the tuned processors exhibit a significant reduction in frequency spread compared to untuned processors. The chip-level global offset observed on tuned chips is likely originating from a combined effect of (1) an operational mismatch between the reserved aging budget for tuning targeting and the actual aging amount until the start of the cooldown, and (2) offset in target resistance assignment due to a systematic error in empirical frequency prediction. However, we emphasize that a global offset of an order of a few percent is typically not a concern for arriving at the desired Hamiltonian pattern on a single-chip quantum processor because it does not affect achieving desired detuning between neighboring qubits for high-fidelity two-qubit gates. To empirically quantify the frequency tuning precision on real quantum processors, in Fig. 5(b), we plot the statistical aggregates of the data shown in Fig. 5 (a) after subtracting the global offset from each tuned and untuned chip. The distributions outlined by Gaussian fit give a standard deviation of 18.4 MHz on tuned and 93.5 MHz on untuned quantum processors, corresponding to 0.40\% and 2.02\% of the average design frequency (4628 MHz). The appearance of symmetric frequency distribution after ABAA tuning also indicates that the tuned spread is independent of the amount of ABAA tuning and the time elapsed after tuning.  

\begin{figure}[t]
\includegraphics[width=\columnwidth]{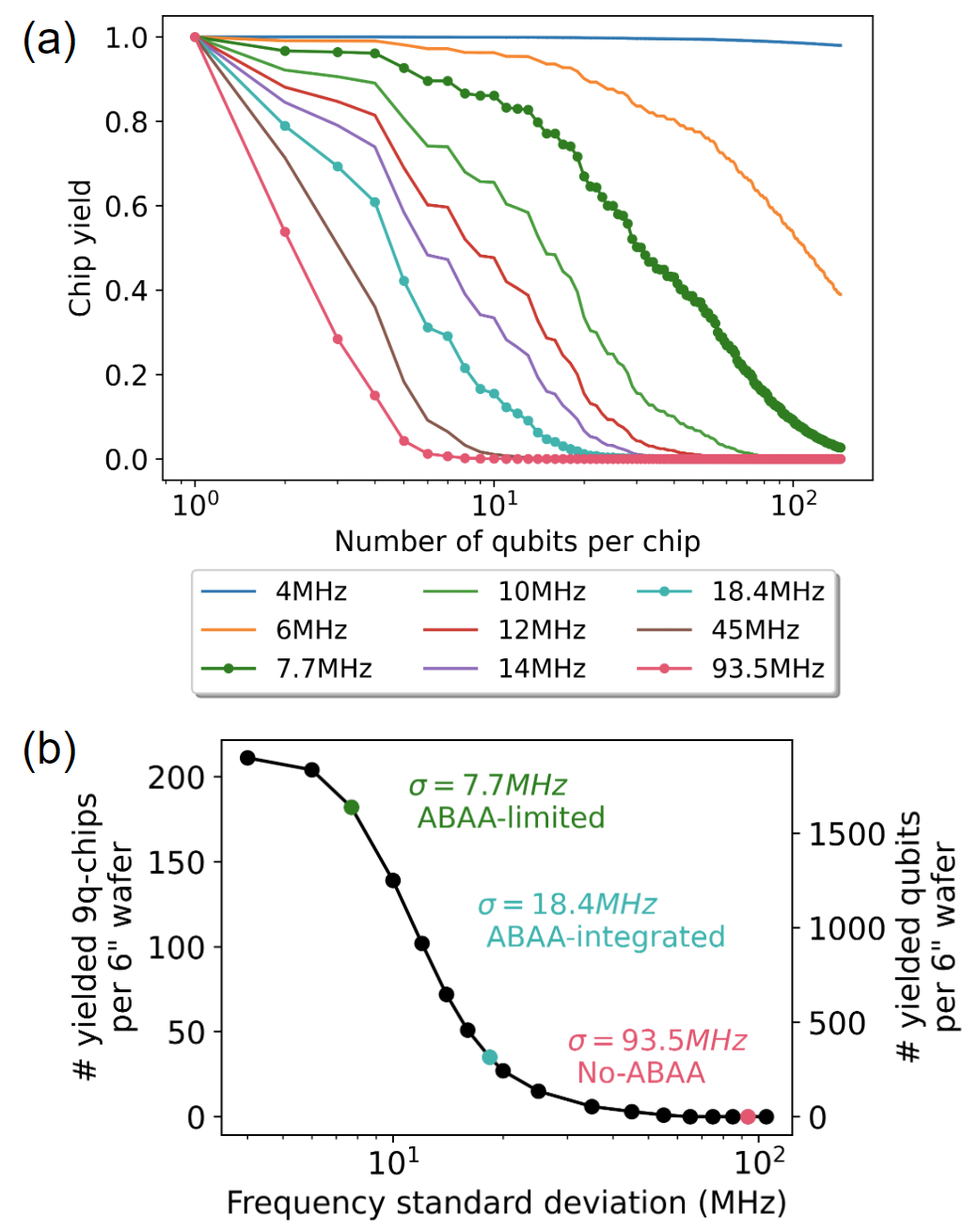}
\caption{Statistical yield analysis as a function of qubit number and frequency spread. Yield here is defined as the probability of all qubit-qubit edges at $f_{01}^{\rm max}$ falling in the 20-130 MHz detuning range, desired for high-fidelity iSWAP gates. (a) Simulated chip yield as a function of the number of qubits per chip over a range of frequency tuning precision. The design Hamiltonian is constructed by tiling a 9-qubit lattice design that ensures all detunings fall within the 40-110 MHz range both within the 9-qubit lattice and along the stitching edges. Curves with markers are relevant to measured quantities - 93.5 MHz is the spread observed in untuned devices, 18.4 MHz is the empirical spread measured in ABAA-tuned devices, and 7.7 MHz is the projected spread in ABAA-tuned devices based on the resistance tuning precision without other sources of uncertainties. (b) Number of yielded, tileable 9-qubit chips (left y-axis) per 6" wafer assuming a total of 212 9-qubit dice per wafer. The right y-axis corresponds to the total number of qubits on the yielded 9-qubit chips per wafer. 
}
\label{Fig:Figure7_v3}
\end{figure}

We characterized two-qubit gate fidelity on ABAA-tuned 9-qubit processors with design Hamiltonians optimized for iSWAP gates. To minimize the impact of flux noise on the dephasing time of the modulated qubit, qubit frequencies are designed such that the nearest neighbor qubit-qubit detuning is centered around $\sim 50~\mathrm{MHz}$. This ensures the modulated qubit will have less frequency excursion from its maximum frequency to bring it into resonance with its neighbors. In addition, the Hamiltonian is designed to make the qubit-qubit detunings large enough that the residual ZZ interaction is minimal at zero-coupling bias. Figure 6 (a) shows the design Hamiltonian of one of the ABAA-tuned 9-qubit processors [tuned \#3 in Fig. 5(a)], where the qubit-qubit frequency detunings range from 20-80 MHz, with a median qubit-qubit detuning of 50.5 MHz. The yellow arrows show which qubit's frequency is modulated among two neighboring qubits in resonance to activate the iSWAP gate. We allocate the qubit frequencies such that there are no more than two edges (or gates) activated by modulating the same qubit. This minimizes the number of frequency collisions between neighboring qubits during gate operation. Figure 6 (b) shows the qubit-qubit detunings that are measured after ABAA at $g=0$ tunable coupler flux bias conditions, with detunings at all edges within the desired small detuning range and a median detuning of 49 MHz, and used for two-qubit gate characterization in Fig. 6 (c). Only one qubit (qubit 0) needs to be parked off of $f_{01}^{\rm max}$ (by -27 MHz) to arrive at the desired frequency configuration shown in Fig. 6 (b), highlighting the excellent frequency allocation after ABAA tuning. Without parking any qubit off of its $f_{01}^{\rm max}$, the difference between the measured and designed detunings is on average -10.8 $\pm$ 22.5 MHz across the lattice, which is consistent with the statistical detuning error of $\pm$ 26.0 MHz that is expected from the measured single-qubit frequency tuning precision of 18.4 MHz. We calibrated parametric resonance two-qubit iSWAP gates \cite{Sete_para_2021} on each edge of the 9-qubit device. Parametric resonance gates are activated by modulating the frequency of the higher-frequency qubits to bring the qubits into resonance. We benchmark the iSWAP gates with interleaved randomized benchmarking (iRB) \cite{Magesan2012}. Figure 6 (c) shows the distribution of the iSWAP iRB fidelities, with fidelity as high as $99.51 \pm 0.20 \%$ and a median fidelity reaching $99.00\%$. The median two-qubit gate time is 68 ns, including a total of 16 ns padding before and after the flux pulse. 

We also characterized a second ABAA-tuned 9-qubit device where we moved two qubits off of $f_{01}^{\rm max}$ to avoid frequency collisions and measured qubit-qubit detunings with a median of 48 MHz, which is very close to the design qubit-qubit detuning. On this device, we characterized parametric resonance iSWAP gates on each edge and benchmarked their performance with iRB, getting a median fidelity of $99.22\%$ and mean fidelity of $99.13 \pm 0.12\%$.

\section{Yield analysis and scaling}
To illustrate the impact of ABAA frequency tuning on yielding tunable qubit lattices at scale, we perform statistical yield modeling analysis as a function of qubit number and frequency tuning precision. We apply the simulation to a scalable Hamiltonian with edge (qubit-qubit) $f_{01}^{\rm max}$ detunings designed in the $40-110\ \mathrm{MHz}$ range and define yield as the probability of all nearest neighbor 2-qubit edges at $f_{01}^{\rm max}$ falling in the $20-130\ \mathrm{MHz}$ detuning range (i.e., desired for high-fidelity iSWAP gates). Here the Hamiltonian is constructed by tiling a 9-qubit (3$\times$3 lattice) unit-cell design that ensures detunings fall within the $40-110\ \mathrm{MHz}$ range within the 9-qubit lattice as well as along stitching edges, allowing 20 MHz buffer zones from the yield boundaries on both ends. This Hamiltonian is scalable because it enables up-scaling quantum processors by stitching yielded 9-qubit chips via inter-module couplers \cite{gold2021entanglement,Field2023}. Figure 7 (a) shows the simulated chip yield as a function of the number of qubits on the chip, over a range of frequency tuning precision. Figure 7 (b) assumes a layout of 212 dice of 9-qubit chips on a 6” wafer and plots the predicted number of yielded 9-qubit chips (left y-axis) and the corresponding number of yielded qubits (right y-axis) per wafer as a function of frequency tuning precision. The three curves with markers in Fig. 7 (a) and the three color-coded data points in Fig. 7 (b) correspond to the relevant tuning precisions demonstrated in this study. (1) With the frequency spread of 93.5 MHz on as-fabricated devices, it is virtually impossible to yield any chip beyond the few-qubit chip scale. The number of yielded 9-qubit chips per wafer is approximately zero. (2) With the empirical frequency tuning precision of 18.4 MHz demonstrated using ABAA in an integrated process flow, there is a $\sim 17\%$ chance for a 9-qubit chip to meet the detuning range while parking at $f_{01}^{\rm max}$. This corresponds to yielding approximately 35 tileable 9-qubit chips (315 yielded qubits total) per 6” wafer. (3) With the frequency-equivalent resistance tuning precision of 7.7 MHz of ABAA, a 9-qubit chip yield of $86\%$ can be achieved, which corresponds to yielding 182 tileable 9-qubit chips per 6” wafer, sufficient for tiling into a scaled-up quantum processor of over 1500 qubits. We also note that the chip yield at 7.7 MHz frequency spread [green curve in Fig. 7 (a)] does not drop to zero till surpassing the 100-qubit chip scale, indicating possibilities of yielding monolithic chips of 100-qubit. These results demonstrate ABAA as a viable post-fabrication frequency tuning technique for achieving high detuning-edge yield beyond the 1000-qubit scale on modular superconducting quantum processors.

\section{Conclusion}
In this work, we characterized the qubit relaxation behavior during ABAA and demonstrated the cutting-edge frequency tuning capability using ABAA. Baseline statistics of hundreds of ABAA-tuned tunable transmon qubits yield a frequency-equivalent resistance tuning precision of 7.7 MHz in $f_{01}^{\rm max}$, sufficient for achieving high detuning-edge-yield beyond the 1000-qubit scale. Cryogenic cooldown measurements of tuned and untuned qubits showed evidence of improved qubit coherence after ABAA with no measurable impact on tunability. Despite a global shift due to calibration offset and aging, we experimentally demonstrated an empirical frequency tuning precision of 18.4 MHz when targeting the designed Hamiltonians on multi-qubit quantum processors. We demonstrated high-fidelity parametric resonance two-qubit iSWAP gates on two ABAA-tuned 9-qubit processors with a median two-qubit fidelity of $99.00\%$ and $99.22\%$, respectively. The results demonstrated in this study prove that the ABAA technique is an easy-to-implement and efficient method for precision frequency tuning and Hamiltonian targeting on scaled-up superconducting quantum processors. Future work includes improving frequency prediction precision and a better understanding of the junction relaxation and aging mechanism to explore the ultimate tuning precision using ABAA, improving control of the global frequency offset at the chip level for improving scaling yield in a modular quantum processor architecture \cite{gold2021entanglement}, and simultaneous tuning of multiple qubits in parallel for throughput improvement. 

\section*{acknowledgments}
We thank the Rigetti Fab-1 team for junction process development and sample fabrication.

\section*{contributions}
X.Wang conceived and implemented the precision freuqency tuning using ABAA and wrote the manuscript; J.H. performed frequency prediction calibration and yield simulation; E.S. designed the iSWAP-optimized Hamiltonian, characterized the two-qubit gates, and ran the data analysis; X.Wu, S.P., G.H., and C.E. performed qubit frequency characterization, and data analysis; C.K. performed qubit coherence characterization; M.F. and J.H. helped develop the measurement apparatus; H.C. and N.S. developed the JJ fabrication process; R.K., J.M., and K.Y. provided important suggestions and leadership; A.B. and D.P.P. are co-PIs who supervised the project. All authors contributed to the writing and editing of the manuscript.

\bibliography{VIA.bib}
\end{document}